\documentstyle[epsfig,12pt]{article}
\begin{document}

\title{A Phenomenological Analysis of Non-resonant Charm Meson Decays}

\author{I. Bediaga, C. G\"obel\\
{\small \it Centro Brasileiro de Pesquisas F\'\i sicas, R. Dr. Xavier Sigaud 150}\\
{\small \it 22290 -- 180 --  Rio de Janeiro, RJ, Brazil}\\
{\small and} \\
R. M\'endez--Galain\\
{\small \it Instituto de F\'{\i}sica, Facultad de Ingenier\'{\i}a,
CC 30}\\ {\small \it CP 11000 Montevideo, Uruguay}}

\maketitle

\begin{abstract}
{\footnotesize We analyse the consequences of the usual assumption of a
constant function to fit non-resonant decays from experimental Dalitz 
plot describing charmed meson decays. We first show, using the $D^+\rightarrow \bar{K}^0\pi^+\pi^0$ 
decay channel as an example, how an inadequate extraction of the 
non-resonant contribution could yield incorrect measurements for the 
resonant channels. We analyse how the correct study of this decay 
will provide a test for the validity of factorization in D meson 
decays. Finally, we show how form factors could be extracted from 
non-resonant decays. We particularly discuss about the form factor 
that can be 
measured from the $D^+_s\rightarrow \pi^-\pi^+\pi^+$ decay. We emphasize on its relevance for the 
study of the decay $\tau \to \nu_{\tau} 3\pi$ and the extraction of 
the $a_1$ meson width.}
\end{abstract}

\section{Introduction}

Many body charm meson decays seem to be largely dominated by intermediate 
resonances. Experimental data have been studied using the 
powerful Dalitz plot technique which brings information on both the 
kinematics and the dynamics of the decay \cite{dalitz}. 

In a $D$ meson three body decay, the intermediate resonant channels
and the direct non-resonant one contribute to the final state.
The Dalitz plot can thus present a complex interference of all these
contributions. To extract them from the plot, one has to use appropriate
fitting functions for each channel.
 Since the discovery of $D$ mesons, data
are fitted using a Breit--Wigner function for each resonance
amplitude \cite{jackson} while the non-resonant (NR) contribution 
has usually been 
considered as a phase space independent, constant function. 

In a recent paper\cite{prl}, we have shown that the last hypothesis 
cannot be safely considered for the study of D meson decays as it proceeds via 
a weak interaction. Indeed, in weak 
interactions at the partonic level
helicity plays a central role; thus one could expect 
the amplitude of the reaction to have important variations within the phase
space. In ref. \cite{prl}, the 
NR contribution to $D^+\rightarrow K^-\pi^+\pi^+$ decay has been evaluated using 
factorization and an effective hamiltonian for the partonic interaction 
\cite{bsw}. According to this 
calculation, the 
NR contribution does have significant variations along the 
phase space of the reaction. 

To extract data from the Dalitz plot, an adequate parametrization of the 
NR contribution is crucial. A correct extraction of the NR 
contribution  could yield important 
information on the physics involving the decay; particularly,
it can bring direct measurements of some form factors. Moreover, using an 
inadequate parametrization for the NR contribution, the whole decay 
pattern could be wrong:
we could be ascribing to a given resonance those variations corresponding 
to the NR part. 

The purpose of this work is to present two examples concerning these ideas. 
First, we will analyse the decay $D^+\rightarrow \bar{K}^0\pi^+\pi^0$ , since its 
$\bar K^*(892)^0\pi^0$ partial decay 
width seems to be too large. Second, we will show how the $D^+_s\rightarrow \pi^-\pi^+\pi^+$ decay is 
particularly well suited to extract a form factor which is relevant in $\tau$ 
and $a_1$-meson physics. 

\section{The  $D^+ \rightarrow \bar K^0 \pi^+ \pi^0$ decay}

The resonant decay $D^+ \rightarrow \bar K^*(892)^0 \pi^+$ has been 
measured in two different ways, according to the detected final 
state: $B(D^+ \to \bar K^{*0} \pi^+) \times B(\bar K^{*0} \to \bar 
K^0 \pi^0)$, which is extracted from the Dalitz plot of the 
decay $D^+\rightarrow \bar{K}^0\pi^+\pi^0$; and $B(D^+ \to \bar K^{*0} \pi^+) \times B(\bar K^{*0} 
\to K^- \pi^+)$, extracted from the decay $D^+\rightarrow K^-\pi^+\pi^+$.  
MarkIII  reported \cite{mark3} an ``apparent discrepancy''
between the two measurements:  
$B(D^+ \to \bar K^{*0} \pi^+) 
= (5.9 \pm 1.9 \pm 2.5)\%$ when the final state is $\bar K^0 
\pi^0 \pi^+$ and 
$B(D^+ \to \bar K^{*0} \pi^+) = (1.8 \pm 0.2 \pm 1.0)\%$ when 
the final state is $K^- \pi^+ \pi^+$.

The last measurement has been confirmed by other
experiments\cite{na14,691,687} while the decay $D^+\rightarrow \bar{K}^0\pi^+\pi^0$ has only been measured by MarkIII.
It is then natural to think on a possible systematic error in the 
extraction of the $\bar K^*(892)^0$ resonance from the $D^+\rightarrow \bar{K}^0\pi^+\pi^0$ 
Dalitz plot. One possibility is that events coming from the NR 
contribution to the decay could have been incorrectly considered 
as originated from the $\bar K^*(892)^0$ resonant channel; thus,
the latter has been artificially enhanced.

To support this hypothesis, we are presenting here a calculation for 
the NR part of the decay $D^+\rightarrow \bar{K}^0\pi^+\pi^0$ which shows up precisely an 
important bump near the $\bar K^*(892)^0$ peak. The calculation 
is based in factorization \cite{fak} and in an effective Hamiltonian 
\cite{bsw,buras} for the partonic interaction as in ref. \cite{prl}.

The effective partonic Hamiltonian is\cite{bsw,buras}

\begin{equation}
{\cal H}_{eff} = \frac{G_F}{\sqrt{2}} \cos ^2 \theta_c [ a_1 
:(\bar{s}c)(\bar{u}d): + a_2 :(\bar{s}d)(\bar{u}c):]
\label{heff}
\end{equation}
where $(\bar{q}q')$ is a short-hand notation for $\bar{q} \gamma^{\mu} 
(1-\gamma_5)q'$. The 
coefficients
$a_1$ and $a_2$ characterize the contribution of the effective charged 
and neutral 
currents respectively,
which include short-distance QCD effects. Figure 1 shows the six diagrams  
contributing to the amplitude
${\cal M}^{NR}_{D^+ \rightarrow \bar K^0 \pi^+\pi^0}$. 
Using factorization one obtains the following decomposition\cite{foot1} for the 
hadronic amplitude of the non-resonant decay
\begin{eqnarray}
{\cal M }^{NR}_{D^+ \to \bar K^0 \pi^0 \pi^+} & = &
 \frac{G_F}{\sqrt{2}} \cos ^2 \theta_c [ a_1 \langle \bar K^0\pi^0|
\bar{s}c|D^+\rangle \langle \pi^+|\bar{u}d|0\rangle 
\nonumber \\
& & + a_2 
\langle \bar K^0|\bar{s}d|0\rangle\langle \pi^0 \pi^+|\bar{u}c|D^+\rangle 
\; \; + \; \; a_1 \langle \bar K^0|\bar{s}c|D^+
\rangle \langle \pi^0 \pi^+|\bar{u}d|0\rangle 
\nonumber \\
& & + a_2 
\langle \bar K^0\pi^0|\bar{s}d|0\rangle\langle \pi^+|\bar{u}c|D^+\rangle ] ~.
\label{mdkpp}
\end{eqnarray}

Following ref. \cite{prl}, the four contributions can be written as
\begin{equation}
\langle \bar K^0\pi^0|\bar{s}c|D^+\rangle \langle \pi^+|\bar{u}d|0\rangle = 
\frac{1}{\sqrt{2}} f_{\pi}   m^2_{\pi} F^{\bar K^0\pi^0}_{4} 
(m_{\pi^0\pi^+}^2, m_{\bar K^0\pi^+}^2) \; \; \; \; ,
\label{diaga}
\end{equation}
\begin{equation}
\langle \pi^0\pi^+|\bar{u}c|D^+\rangle \langle \bar K^0|\bar{s}d|0\rangle = 
\frac{2}{\sqrt{2}} f_{K}   m^2_{\bar K^0} F^{\pi^+\pi^0}_{4} 
 (m_{\bar K^0\pi^+}^2,  m_{\bar K^0\pi^0}^2)\; \; \; \; ,
\label{diagb}
\end{equation}
\begin{eqnarray}
& & 
\langle \bar K^0|\bar{s}c|D^+\rangle \langle \pi^0 \pi^+|
\bar{u}d|0\rangle 
= \frac{2}{\sqrt{2}} 
%\left
\; \{ \; 
F^{1}_{DK}(m_{\pi^0\pi^+}^2) f^+_{\pi^0\pi^+}
(m_{\pi^0\pi^+}^2) 
\: (m_{\bar K^0 \pi^0}^2  - m_{\bar K^0\pi^+}^2) \; \; + 
%\right. \}
\nonumber \\
& & 
%\left. \{
[ F^{1}_{DK}(m_{\pi^0\pi^+}^2) f^+_{\pi^0\pi^+} 
(m_{\pi^0\pi^+}^2)  - 
F^{0}_{DK}(m_{\pi^0\pi^+}^2) f^0_{\pi^0\pi^+} (m_{\pi^0\pi^+}^2) 
 ] \:  \frac{(m_D^2 - m_{\bar K^0}^2)(m_{\pi^+}^2 - m_{\pi^0}^2)}
{m_{\pi^0\pi^+}^2} 
%\right
\; \} 
\label{diagc}
\end{eqnarray}
and 
\begin{eqnarray}
& & 
\langle \pi^+|\bar{u}c|D^+\rangle \langle \bar K^0 \pi^0|
\bar{s}d|0\rangle = 
\frac{1}{\sqrt{2}} 
%\left
\; \{ \; F^{1}_{D\pi}(m_{\bar K^0\pi^0}^2) f^+_{K\pi}
(m_{\bar K^0\pi^0}^2) 
\: (m_{\pi^0 \pi^+}^2  - m_{\bar K^0\pi^+}^2) \; \; + 
%\right. \}
\nonumber \\
& &
%\left. \{
[ F^{1}_{D\pi}(m_{\bar K^0\pi^0}^2) f^+_{K\pi} 
(m_{\bar K^0\pi^0}^2)  - 
F^{0}_{D\pi}(m_{\bar K^0\pi^0}^2) f^0_{K\pi} (m_{\bar K^0\pi^0}^2) 
]\:  \frac{(m_D^2 - m_{\pi^+}^2)(m_{\bar K^0}^2 - m_{\pi^0}^2)}
{m_{\bar K^0\pi^0}^2} 
%\right
\; \} 
\label{diagd}
\end{eqnarray}

We have introduced the 3 invariants $m_{\bar K^0\pi^+}^2 \equiv (p_{\bar K^0} 
+ p_{\pi^+})^2$,
$m_{\bar K^0\pi^0}^2 \equiv (p_{\bar K^0} + p_{\pi^0})^2$ and 
$m_{\pi^0\pi^+}^2 \equiv (p_{\pi^0} + p_{\pi^+})^2$ and use has been 
made of the identity 
\begin{equation}
m_{D^+}^2 + m_{\bar K^0}^2 + m_{\pi^+}^2 + m_{\pi^0}^2 = 
m_{\bar K^0\pi^+}^2 + m_{\bar K^0\pi^0}^2 + m_{\pi^0\pi^+}^2 \; \;. 
\label{ident}
\end{equation}
The $1/\sqrt{2}$ factors come from the $\pi^0$ wave functions.
The ten form factors originate from the hadronic matrix elements\cite{prl};
we will be back to them later on. 

 Diagrams (a), (b) and (e) of Fig. 
1 exhibit an 
external light pseudoscalar meson (P). This yields a contribution 
proportional to $f_P m_P^2$ as one can see from eqs. (\ref{diaga}) 
and (\ref{diagb}). The other diagrams, i.e., (c), (d) and (f), 
produce contributions proportional to $m_D^2$
as one can see from eqs. (\ref{diagc}) and (\ref{diagd}), together with
 eq. (\ref{ident}). Thus the two first contributions in eq. (\ref{mdkpp}) 
 can be safely neglected in favor of the last two. Moreover, the second term
in eq. (\ref{diagc}) can be neglected as it is proportional
to $(m^2_{\pi^+} - m^2_{\pi^0})$.

The eight form factors entering in eqs. (\ref{diagc}) and (\ref{diagd})
are written as
\begin{equation}
F^J_{DP} (q^2) = \frac{F^J_{DP} (0)}{(1-q^2/{m^2_{DP,J}})}
\label{ffDP}
\end{equation}
where $J=0$ or 1 and $P=K$ or $\pi$, and
\begin{equation}
f^i_{AB} (q^2) = f^i_{AB} (0) \; (1 + \lambda^i_{AB} q^2/m_{\pi}^2)
\label{ff+0}
\end{equation}
where $i=+$ or 0 and $AB= K \pi$ or $\pi^+\pi^0$.
According to the remark above, only six of them contribute to our calculation.
Five of these form factors have been either measured from semileptonic decays, 
or calculated using lattice QCD or different quark models. There are no 
major discrepancies in the
literature \cite{UKQCD}:  $F^1_{D\bar K^0}
(0) = 0.75 \pm 0.1$, $m^1_{D\bar K^0} = 2.0 \pm 0.2$; $F^1_{D\pi}
(0) = 0.75 \pm 0.15$, $m^1_{D\pi} = 2.1 \pm 0.2$; $F^0_{D\pi}
(0) = 0.75 \pm 0.15$, $m^0_{D\pi} = 2.2 \pm 0.2$; $f^+_{\bar K^0\pi^0}(0)
= 0.7 \pm 0.1$, $\lambda^+_{\bar K^0\pi^0} = 0.028 \pm 0.002$; 
$f^0_{\bar K^0\pi^0}(0) =
0.7 \pm 0.1$, $\lambda^0_{\bar K^0\pi^0} = 0.004 \pm 0.007$. The sixth
form factor entering in our calculation, $f^+_{\pi^+\pi^0} (q^2)$, has neither 
been measured nor obtained
using lattice calculations. One can only hint $f^+_{\pi^+\pi^0}(0)$ calculating
\cite{okun} the
decay width $\pi^+ \to \pi^0 e^+ \nu_e$ and comparing it with experiment, 
to get $f^+_{\pi^+\pi^0}(0)$ = 1.4. 

Finally, the only measurement we have for the effective parameters $a_1$ and 
$a_2$ come from the fit of two body charm meson decays. It has been found
\cite{a1a2} $a_1 = 1.26 \pm 0.10$ and $a_2 = -0.51 \pm 0.10$. 

We have performed a Monte Carlo simulation of the NR contribution to the decay 
$D^+\rightarrow \bar{K}^0\pi^+\pi^0$ using eqs. (\ref{mdkpp}), (\ref{diagc}) and (\ref{diagd}). Figure 2
presents the Dalitz plot corresponding to the NR contribution to the
decay, according to the calculation presented above. Figure 3 shows the density 
of events as a function of the
invariant variable $m^2_{\bar K^0 \pi^0}$. It presents a pronounced bump 
centered in $m^2_{\bar K^0 \pi^0} \approx 0.65$ GeV$^2$. The figures shown 
have been obtained using
the central values of all the parameters presented above. The bump remains unchanged 
doing any variation of these parameters within the region allowed by experiment
and lattice calculation and a very large variations for the unknown
parameters defining the $f^+_{\pi^+\pi^0} (q^2)$ form factor. The bump is also
unchanged within a large variation of the ratio $a_2/a_1$. 

This strong stability of the bump in the simulation results shown in Fig. 3 
is due to the following:
it turns out that if the ratio $|a_2/a_1|$ is not very large (in fact, smaller 
than about 2.5), then the contribution of eq. (\ref{diagc}) largely drives the 
behavior of the $m^2_{\bar K^0 \pi^0}$ distribution. Thus, we can write
\begin{equation}
{\cal M }^{NR}_{D^+ \to \bar K^0 \pi^0 \pi^+} \propto
F^{1^-}_{DK}(m_{\pi^0\pi^+}^2) f^+_{\pi^0\pi^+}(m_{\pi^0\pi^+}^2) 
\: (m_{\bar K^0 \pi^0}^2  - m_{\bar K^0\pi^+}^2)
\label{mfin}
\end{equation}
Since the form factors in (\ref{mfin}) depend only on $m_{\pi^0 \pi^+}^2$, 
the $m_{\bar K^0 \pi^0}^2$  distribution of the events is thus almost
independent of the various poorly known quantities associated with the decay.
For comparison, we present in Figure 4 the $m_{\bar K^0 \pi^0}^2$ distribution 
of events when no dynamics is assumed for the NR decay, i.e., 
${\cal M}^{NR}$ = const. The bump in Figure 3 is thus a robust signature 
of a factorization-based calculation.

However, as some non-perturbative QCD effects -- as final state interactions 
and soft gluon exchange -- have been neglected, it is possible that factorization 
does not
suffice to describe the decay. In the extreme case where the non-perturbative effects dominate the decay, the structure predicted above can be washed out. Thus, the experimental determination of a bump 
in the NR contribution
to the decay $D^+\rightarrow \bar{K}^0\pi^+\pi^0$ centered at $m_{\bar K^0 \pi^0}^2 \approx 0.65$ GeV$^2$ 
would be a test for the validity of factorization in $D$ decays.

The bump predicted by this calculation lies near the peak one expects for
the Breit-Wigner distribution corresponding to the $\bar K^*(892)^0$ resonance.
If non-factorizable terms does not completely eliminate the bump, 
many events originated from the NR decay $D^+\rightarrow \bar{K}^0\pi^+\pi^0$ have been probably incorrectly 
ascribed to the $D^+ \to \bar K^*(892)^0 \pi^+$ resonant channel.

For completeness, we present in Figure 5 the $m^2_{K^-\pi^+}$ distribution 
of events for the NR part of the decay $D^+\rightarrow K^-\pi^+\pi^+$. We use the amplitude we 
obtained in ref. \cite{prl}. There is no bump near the the 
$\bar K^*(892)^0$ squared mass. It looks more like a simple phase 
space distribution, as that of Figure 4. 

Thus, the difference between the NR contributions of the decays
$D^+\rightarrow K^-\pi^+\pi^+$ and $D^+\rightarrow \bar{K}^0\pi^+\pi^0$ could explain the different values reported
for the decay $D^+ \to K^*(892)^0 \pi^+$ according to the final state.
It is an example of the indirect consequences of assuming inadequately
a constant NR function to fit data.

\section {The $D^+_s\rightarrow \pi^-\pi^+\pi^+$ decay}

A correct extraction of the NR contribution to a given charmed many body decay
could also have other important advantages. As we have shown above, the NR 
contribution to a given heavy meson decay is written in terms
of various form factors. Thus, its correct extraction from the Dalitz plot
could also be a way to measure those form factors within the whole phase space 
of the reaction. 

Two problems arise here. First, one has to accept that non-factorizable 
effects are small, so that the expression of the NR amplitude in terms of 
the form factors can be simply obtained using factorization hypothesis --- as 
we did above and in ref. \cite{prl}.
Second, even assuming the validity of factorization, those expressions are 
products of
form factors and it is thus complicated to extract separately each of them.

We present here an example in which these two problems are supposed to
be not important. It
is the case of the decay $D^+_s\rightarrow \pi^-\pi^+\pi^+$ . 
The main reason is that there
is only one diagram contributing to the decay and it is an annihilation diagram.
It is shown in Fig. 6. In this case, following Bjorken ideas\cite{bjorken}, factorization looks natural. One thus
expects the decay to be simply described by:
\begin{equation}
{\cal M }^{NR}_{D^+_s \to \pi^+ \pi^+ \pi^-} = \frac{G_F}{\sqrt{2}} \cos ^2 
\theta_c a_1
\langle 0 | A^{\mu} | D^+_s \rangle
\langle \pi^+ \pi^+ \pi^-|A_{\mu}|0\rangle 
\label{ds}
\end{equation}
where
\begin{equation}
\langle 0|A^{\mu}|D^+_s\rangle = - i f_{D_s} p^{\mu}_{D}
\label{dscurr}
\end{equation}
and the second matrix element can be decomposed in 
four form factors\cite{kuhn}. The only remaining term is the axial spin 0 one,
i.e., similar to the one appearing in eqs. (\ref{diaga}) and (\ref{diagb}). 
One obtains \cite{gourdin}
\begin{equation}
{\cal M }_{D^+_s \to \pi^+ \pi^+ \pi^-} = - i \frac{G_F}{\sqrt{2}} \cos ^2 
\theta_c a_1 m_{D_s}^2 f_{D_s} F_4 (m_{\pi^-\pi_1^+}^2,m_{\pi^-\pi_2^+}^2)
\label{dsfin}
\end{equation}
where $m_{\pi^-\pi_1^+}^2 \equiv (p_{\pi^-} + p_{\pi_1^+})^2$ and
$m_{\pi^-\pi_2^+}^2 \equiv 
(p_{\pi^-} + p_{\pi_2^+})^2$. 

The second problem raised above is naturally solved in this particular 
decay: the amplitude is proportional to just one form factor; thus, one
can directly extract it from the plot.

In two body decays of D mesons, amplitudes proportional to $m_D^2$ only happen
through spectator diagrams while those contributions coming from non spectator
diagrams --- as the one we are considering here --- are proportional to the masses
of the final state mesons, and thus less important. Since the amplitude of 
eq. (\ref{dsfin}) is proportional to $m_{D_s}^2$, in principle, it is not small. 
Nevertheless, if one 
assumes PCAC to be valid -- due to the fact that final 
state quarks are light -- one expects this decay
to be small, and this can only happen if the $F_4$ form factor is negligible.
However, the validity of PCAC in this context is not clear, as we will see in 
the following.

The $F_4$ form factor\cite{foot2} has never been measured and there are no 
clear 
theoretical predictions for it. Some authors\cite{weinberg,cpt} proposed 
expressions based in models that are valid only for small values of the
squared momentum transfered to the three pions, $q^2$.
However, in the decay $D^+_s\rightarrow \pi^-\pi^+\pi^+$, $q^2=m_{D_s}^2$.

The measurement of $F_4$ will have important consequences on the understanding 
of $\tau$ and $a_1$ meson physics. The $a_1$ width can be measured through the 
decay $\tau \to \nu_{\tau} 3\pi$ but its value turns out to be 2 or even 3 times 
larger than the value extracted from other measurements\cite{pdg}. The value of 
the $a_1$ width extracted from the decay $\tau \to \nu_{\tau} 3\pi$ strongly 
depends on the magnitude of a possible non-resonant decay, which is driven by 
$F_4$, i.e. the same form factor involved in the NR decay $D^+_s\rightarrow \pi^-\pi^+\pi^+$.

Experimental measurements\cite{opal} of the channel $\tau \to \nu_{\tau} 3\pi$ 
cannot distinguish between models predicting a large amount of PCAC 
breaking\cite{isgur}, i.e. a large $F_4$, from those predicting a small amount 
of this breaking\cite{ks}.
Both a large and a small $F_4$ are acceptable, but the extracted
values of the $a_1$ width can vary by as much as a factor of two when fitting 
data using the first or the second kind of model.

Thus, the correct extraction of the NR part of the decay $D^+_s\rightarrow \pi^-\pi^+\pi^+$ can
bring a first measurement of the form factor $F_4$, clarifying the
amount of PCAC breaking and then helping to extract the correct
value of the $a_1$ meson width. At present, the existent measurements of
the decay $D^+_s\rightarrow \pi^-\pi^+\pi^+$ are not consistent: The branching ratio (BR) for the NR decay
measured from the E691 experiment \cite{691ds} is $1.04 \pm 0.4$ \% -- implying 
a large $F_4$ -- while preliminary results from E687\cite{687ds} give
a value of this BR as small as $0.121 \pm 0.115$ \% -- presenting a smaller
$F_4$.  They have both been obtained using a constant
function to fit the NR contribution.

\section{Summary and conclusions}

In this work, we discuss on some of the consequences of our previous claim\cite{prl}
that NR contributions in $D$ meson decays cannot be fitted with a
constant as they usually are.
We show here that important physics information is hidden in this
contribution which has been loosely considered up to now.
We present two examples to show the information one could obtain
if the NR contribution were correctly extracted from the Dalitz plot.

First, we argue that in the decay $D^+\rightarrow \bar{K}^0\pi^+\pi^0$ events produced via the
 NR channel could have been assumed to be originated from the $\bar K^*(892)^0$
resonant contribution. Using a model based in factorization, we showed that NR 
have a bump near the $\bar K^*(892)^0$ squared mass. This bump is very stable 
within a large variation of some poorly known quantities entering in the 
calculated amplitude. It is thus a strong prediction of factorization.

Second, we claim here that an adequate extraction of the NR contribution
from data could allow us to measure unknown form factors.
The $D^+_s\rightarrow \pi^-\pi^+\pi^+$ is particularly interesting:
its amplitude can be factorized to give a contribution proportional
to the $F_4$ form factor. This form factor drives the spin zero part of the 
axial current matrix element describing the decay in three pions. 
It is very relevant to the decay $\tau \to \nu_{\tau} 3\pi$. Different model 
predictions could be tested and
the longstanding problem concerning the $a_1$ meson width will benefit
from this crucial information.

Coming experiments on charmed meson decays are expected to measure $D^+\rightarrow \bar{K}^0\pi^+\pi^0$ and $D^+_s\rightarrow \pi^-\pi^+\pi^+$
decays with high statistics.
Using a non-constant function for the NR contribution
when fitting the decay from its Dalitz plot, it will be possible to
 extract
adequately the NR contributions. This will a) clarify the eventual discrepancy 
in the $D^+ \rightarrow \bar K^*(892)^0 \pi^+$ decay width, b) test the validity 
of factorization technique when applied to D meson decays and c) bring a first 
measurement of the relevant form factor $F_4$.

\vskip 1 cm

One of us (RMG) would like to thank the Fermilab Theoretical Physics Department,
where part of this work has been done. Two of us (IB and CG) want to thank the 
CNPq (Brazil) for financial support.

\begin{figure}
\begin{center}
\mbox{\psfig{figure=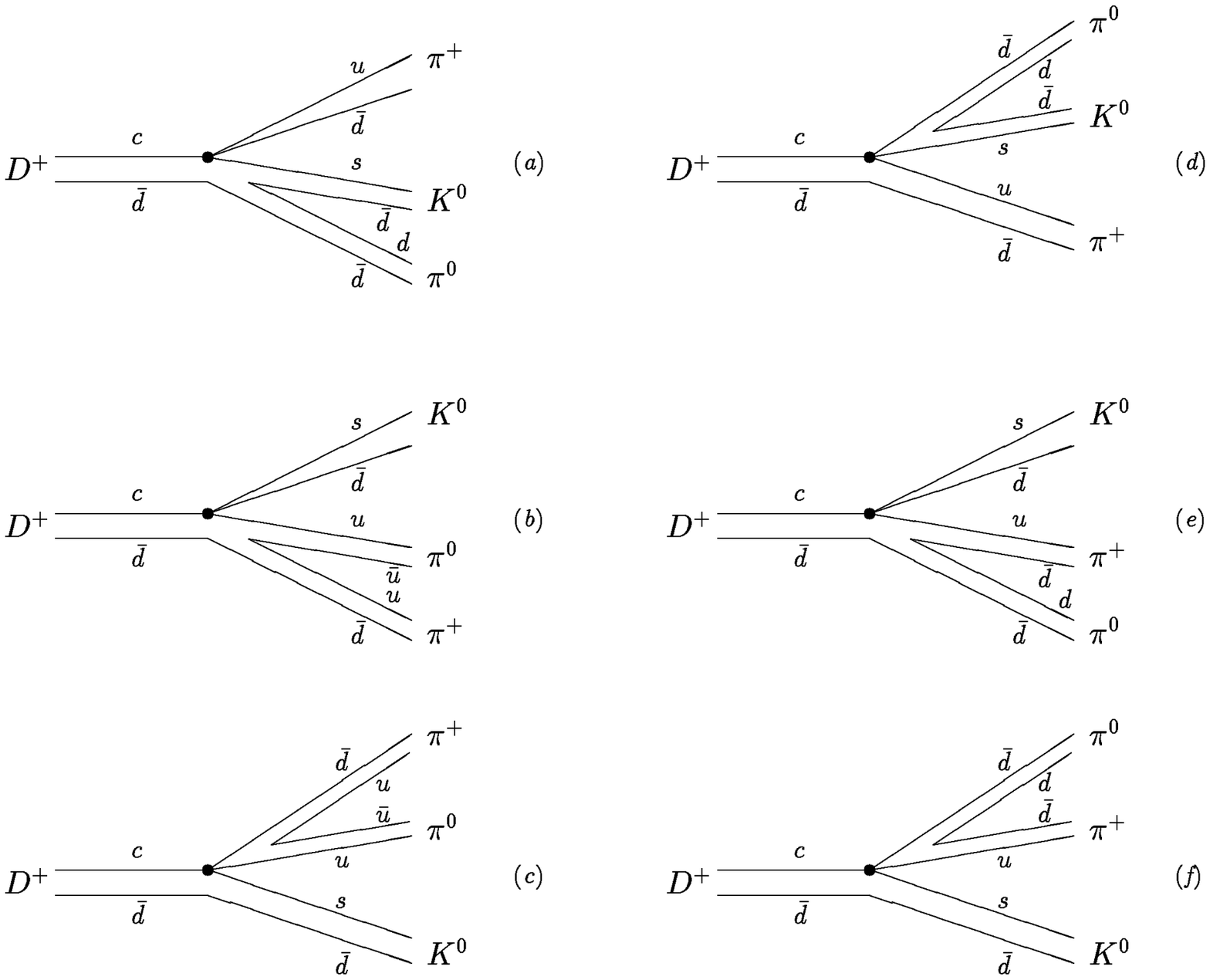,width=16cm}}
\caption{The six diagrams contributing to the decay $D^+ \to \bar K^0 
\pi^0 \pi^+$
according to the effective Hamiltonian of equation (1).}
\end{center}
\protect\label{diagrm}
\end{figure}

\begin{figure}
\begin{center}
\mbox{\psfig{figure=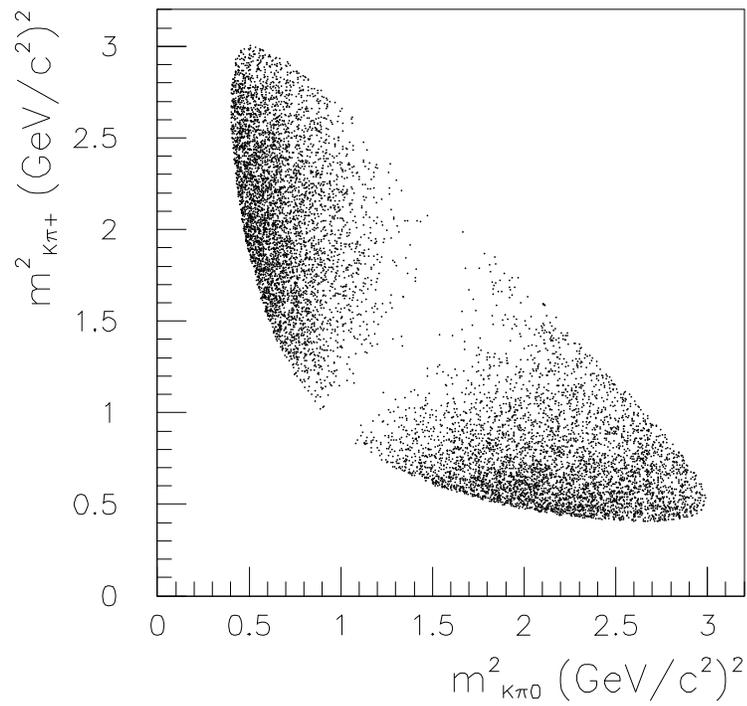,width=13cm}}
\caption{The Dalitz plot for the NR decay $D^+\rightarrow \bar{K}^0\pi^+\pi^0$. It has been obtained with 
Monte 
Carlo simulation weighted by
$|{\cal M}_{D^+ \to K^0 \pi^+ \pi^0}|^2$ as in equations (2), (5) and (6).}
\end{center}
\protect\label{dp}
\end{figure}

\begin{figure}
\begin{center}
\mbox{\psfig{figure=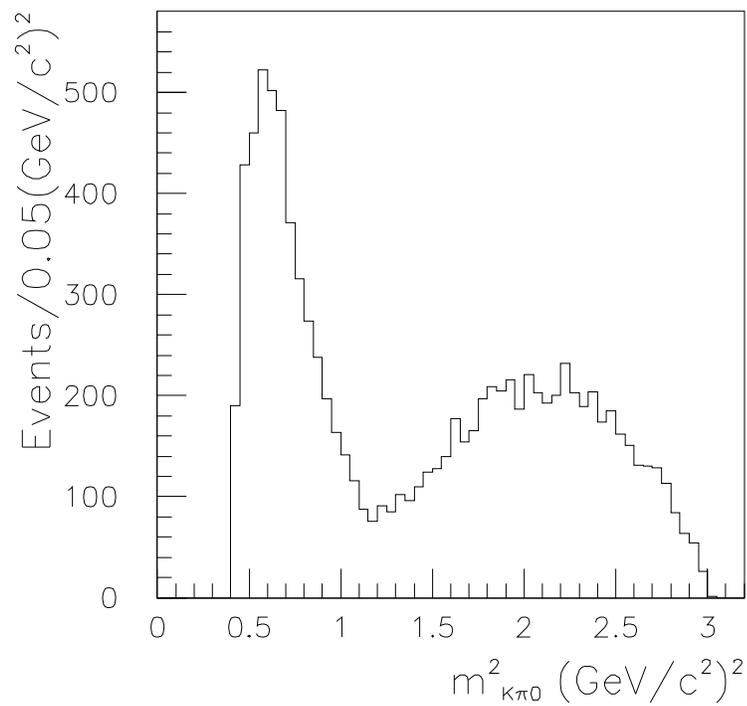,width=13cm}}
\caption{The $m_{\bar K^0\pi^0}^2$ density distribution for the NR decay $D^+\rightarrow \bar{K}^0\pi^+\pi^0$. 
It has been obtained with Monte Carlo simulation weighted by
$|{\cal M}_{D^+ \to K^0 \pi^+ \pi^0}|^2$ as in equations (2), (5) and (6).}
\end{center}
\protect\label{pico}
\end{figure}

\begin{figure}
\begin{center}
\mbox{\psfig{figure=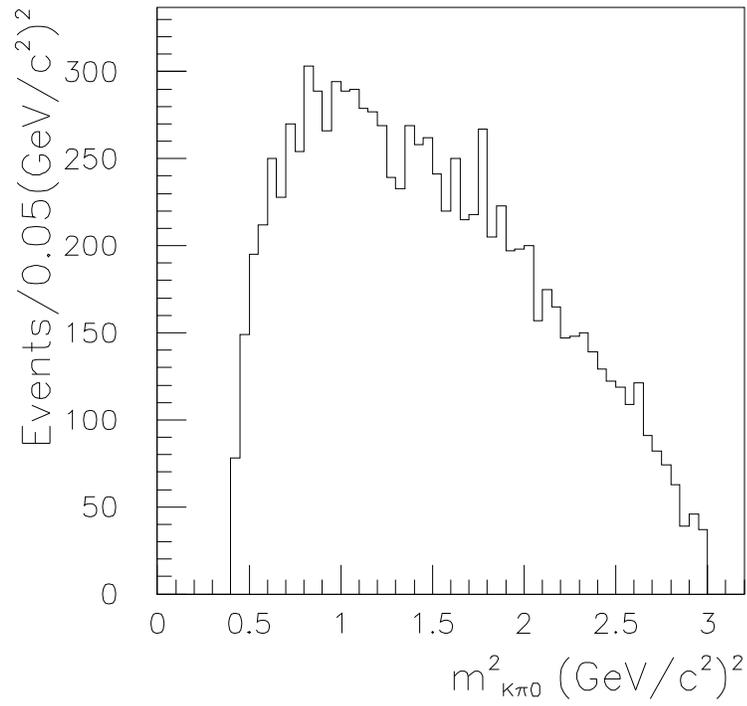,width=13cm}}
\caption{Similar as Fig. 2, but for a flat decay, i.e.,  $|{\cal M}|^2$ = const.}
\end{center}
\protect\label{phsp}
\end{figure}

\begin{figure}
\begin{center}
\mbox{\psfig{figure=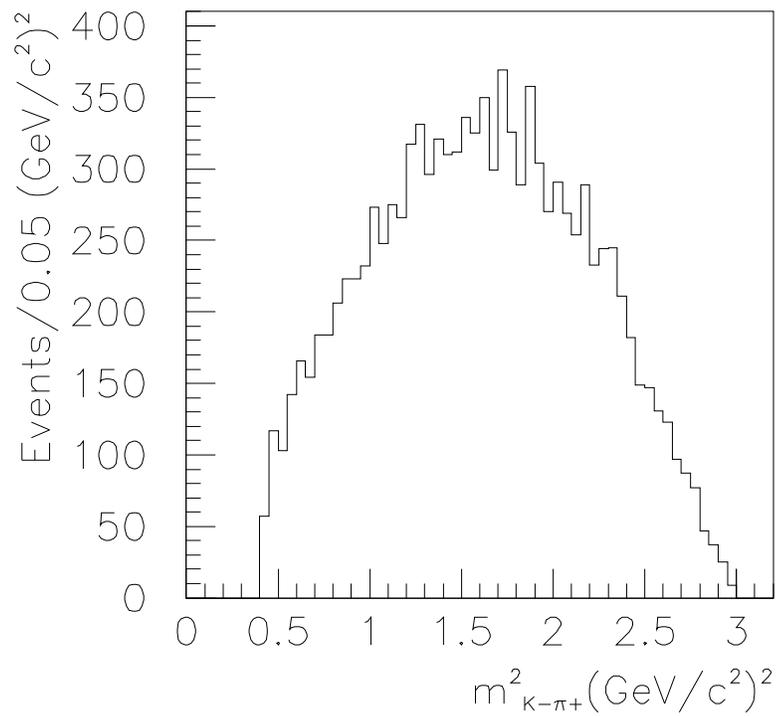,width=13cm}}
\caption{Similar as Fig. 2, but for the decay $D^+ \to K^-\pi^+\pi^+$, using 
the model calculation developed in ref. [3].}
\end{center}
\protect\label{kpipi}
\end{figure}

\begin{figure}
\begin{center}
\mbox{\psfig{figure=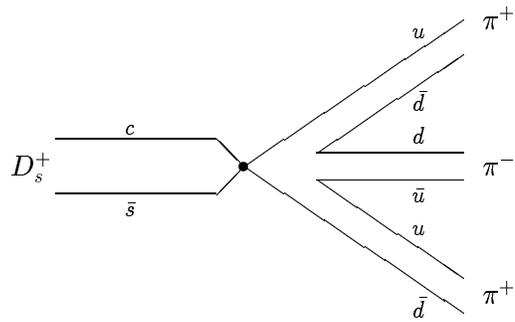,width=16cm}}
\caption{The annihilation diagram dominating the $D^+_s\rightarrow \pi^-\pi^+\pi^+$ decay.}
\end{center}
\protect\label{diagrm2}
\end{figure}


\begin{thebibliography}{999}
\bibitem{foot1} Contributions of the 
two diagrams (b) and (e) are identical;
same for those of diagrams (c) and (f); this explains the factor of 2 in eqs.
(\ref{diagb}) and (\ref{diagc}).
\bibitem{foot2} $F_4$ is also known in the literature as $F_s$, for scalar.
\end{thebibliography}
\end{document}